\documentclass[useAMS,usenatbib]{mn2e}

\usepackage{lmodern}
\usepackage[T1]{fontenc}
\usepackage[pdftex]{color,graphicx}
\usepackage{aas_macros}
\usepackage{amsmath,amsfonts,amssymb}

\renewcommand{\vec}[1]{\ensuremath{\boldsymbol{#1}}}
\newcommand{\der}[2]{\ensuremath{\frac{\mathrm{d}{#1}}{\mathrm{d}{#2}}}}
\newcommand{\dder}[2]{\ensuremath{\frac{\mathrm{d^2}{#1}}{\mathrm{d}{#2}^2}}}
\newcommand{\pder}[2]{\ensuremath{\frac{\partial{#1}}{\partial{#2}}}}
\newcommand{\rmi}{\ensuremath{\mathrm{i}}}
\newcommand{\rme}{\ensuremath{\mathrm{e}}}
\newcommand{\rmd}{\ensuremath{\mathrm{d}}}
\newcommand{\zaver}[1]{\ensuremath{\left<{#1}\right>_z}}

\title[Excitation of SD waves: Simulations]
{The excitation of spiral density waves through turbulent
fluctuations in accretion discs II: Numerical Simulations with MRI driven
turbulence}
\author[T.~Heinemann and J.~C.~B.~Papaloizou]
{T.~Heinemann$^1$ and J.~C.~B.~Papaloizou$^1$\\
$^1$University of Cambridge, Wilberforce Road, Cambridge CB3 0WA}

\date{\today}

\pagerange{\pageref{firstpage}--\pageref{lastpage}} \pubyear{2008}

\begin{document}

\label{firstpage}

\maketitle

\begin{abstract}
We present fully three-dimensional local simulations of compressible MRI
turbulence with the object of studying and elucidating the excitation of the
non-axisymmetric spiral density waves that are observed to always be
present in such simulations.
They are potentially important for affecting protoplanetary migration
through the action of associated stochastic gravitational forces and producing
residual transport in MHD inactive regions through which they may propagate.
The simulations we perform are with zero net flux and produce mean activity
levels corresponding to the Shakura \& Sunyaev
\mbox{$\alpha{}\sim{}5\times{}10^{-3}$}, being at the lower end of the range
usually considered in accretion disc modelling. We reveal the nature of the
mechanism responsible for the excitation of these waves by determining the
time dependent evolution of the Fourier transforms of the participating state
variables. The dominant waves are found to have no vertical structure and to
be excited during periodically repeating swings in which they change from
leading to trailing. The initial phase of the evolution of such a swing is
found to be in excellent agreement with that expected from the WKBJ theory
developed in a preceding paper by Heinemann \& Papaloizou. However, shortly
after the attainment of the expected maximum wave amplitude, the waves begin
to be damped on account of the formation of weak shocks. As expected from the
theory the waves are seen to shorten in radial wavelength as they propagate.
This feature enables nonlinear dissipation to continue in spite of amplitude
decrease. As a consequence the waves are almost always seen to be in the non
linear regime.

We demonstrate that the important source terms causing excitation of the waves
are related to a quantity that reduces to the potential vorticity for small
perturbations from the background state with no vertical dependence. We find
that the root mean square density fluctuations associated with the waves are
positively correlated with both this quantity and the general level of
hydromagnetic turbulence. The mean angular momentum transport associated with
spiral density waves generated in our simulations is
estimated to be a significant fraction of that
associated with the turbulent Reynolds stress.
\end{abstract}


\section{Introduction}\label{sec:introduction}
This is the second of two papers which study the excitation of
spiral density (SD) waves in accretion discs occurring through the
action of stochastic forcing due to turbulence. In practice this has been
taken to arise from the magneto-rotational instability (MRI) but excitation
resulting from stochastic forcing produced by other mechanisms is expected to
lead to similar results. In the preceding paper
\citep[][paper I]{2008arXiv0812.2068H}
we developed a WKBJ theory of the excitation process.
In this paper we study the wave excitation process directly as it is
manifested in fully nonlinear three dimensional numerical simulations, with
the object of elucidating it further and comparing the results with the
predictions of the WKBJ theory developed in paper~I.

The outline of this paper is as follows. In Section~\ref{sec:preliminaries} we
give a brief review of the shearing box model, the basic equations solved in
the simulation, as well as the equations governing the dynamics of SD waves.
We also describe the procedure for performing the Fourier transforms of the
state variables participating in the excited waves in shearing coordinates.
This allows us to identify and then follow the evolution of wave
amplitudes in order to make a detailed comparison with the WKBJ theory
presented in paper~I.

In Section~\ref{sec:simulations}, after briefly describing the simulation set
up and the numerical method used, we present a fully three-dimensional
simulation of compressible MRI turbulence in which we directly observe the
excitation of large scale, non-axisymmetric SD waves. We discuss their main
characteristics and the correlation of the root mean square density
fluctuations associated with them with the level of turbulent activity. We go
on to study the source terms driving the wave excitation and motivate the
simplifying assumption that those
proportional to a quantity that is related to the
potential vorticity dominate.
This assumption enabled us to
derive the linear theory of wave excitation presented in paper~I.  A detailed
comparison between this linear theory and simulation outcomes is presented in
Section~\ref{sec:comparison}. There we follow the time dependent evolution of
the appropriate Fourier amplitudes through a swing cycle, during which the
wave excitation occurs, in detail. We obtain very good agreement with
predictions from the theory developed in paper~I. Finally, we discuss and
summarize our results in Section~\ref{sec:discussion}, indicating the
magnitude and the scaling of the wave angular momentum flux with the Reynolds
stress and also the wavelength in the azimuthal direction for which the wave
excitation phenomenon is optimal.

\section{Preliminaries}\label{sec:preliminaries}

\subsection{The shearing box model}
As in paper~I we consider an isothermal conducting gas in the shearing box
approximation in the absence of vertical stratification. The Cartesian
coordinate system $(x,y,z)$ is such that $x$ corresponds to radius, $y$ to
azimuth, and $z$ denotes the vertical direction. The equations of motions
consist of the continuity equation
\begin{equation}\label{eq:continuity}
\mathcal{D}\rho + \nabla\cdot\vec{p} = 0,
\end{equation}
the momentum equation
\begin{equation}\label{eq:momentum}
\mathcal{D}\vec{p} = -c^2\nabla\rho - 2\vec{\Omega}\times\vec{p}
+ q\Omega p_x\vec{e}_y + \nabla\cdot\vec{T}
+ \nabla\cdot(2\rho\nu\vec{S})
\end{equation}
and the induction equation
\begin{equation}\label{eq:induction}
\mathcal{D}\vec{B} =
\nabla\times(\vec{u}\times\vec{B} - \eta\nabla\times\vec{B})
- q\Omega B_x\vec{e}_y
\end{equation}
where the differential operator
\begin{equation}\label{eq:Dshear}
\mathcal{D} = \partial_t - q\Omega x\partial_y,
\end{equation}
accounts for advection by the linear shear $-q\Omega{}x\vec{e}_y$,
$\vec{\Omega}=\Omega\vec{e}_z$ is the angular velocity, $\rho$ is the
density, the linear momentum density $\vec{p}=\rho\vec{u}$ defined in terms of
the velocity deviations $\vec{u}=(u_x,u_y,u_z)$ from the mean shear,
$c=H\Omega$ is the isothermal sound speed, $H$ is the nominal density scale
height, the magnetic field $\vec{B}=(B_x,B_y,B_z)$ is normalised by the square
root of the vacuum permeability $\mu_0$, in which case the nonlinear stress
tensor $\vec{T}$ has components
\begin{equation}
\label{eq:stress-tensor}
T_{ij} = B_i B_j - \delta_{ij}\vec{B}^2/2 - \rho u_i u_j.
\end{equation}
$\vec{S}$ is the traceless rate-of-strain tensor whose components are given
by
\begin{equation*}
S_{ij} = \frac{1}{2}\left(\pder{u_i}{x_j} + \pder{u_j}{x_i}
- \frac{2}{3}\delta_{ij}\nabla\cdot\vec{u}\right)
- \frac{q\Omega}{2}\left(\delta_{ix}\delta_{jy}+\delta_{iy}\delta_{jx}\right)
\end{equation*}
which are seen to include terms arising due to the linear shear.
The kinematic viscosity is $\nu$ and the resistivity is $\eta$.

We consider the governing equations (\ref{eq:continuity}--\ref{eq:induction})
to be subject to `shearing periodic' periodic boundary conditions, i.e.
\begin{subequations}\label{eq:shearing-periodic-bc}
\begin{equation}\label{eq:shearing-periodic-bcx}
f(x + L_x,y - q\Omega t L_x,z,t) = f(x,y,z,t),
\end{equation}
\begin{equation}\label{eq:shearing-periodic-bcy}
f(x,y + L_y,z,t) = f(x,y,z,t)
\end{equation}
and
\begin{equation}\label{eq:shearing-periodic-bcz}
f(x,y,z + L_z,t) = f(x,y,z,t).
\end{equation}
\end{subequations}
From (\ref{eq:shearing-periodic-bc}) we see that the domain becomes fully
periodic once every shearing time
\begin{equation}\label{eq:shearing-time}
\delta T_\mathrm{s} = \frac{L_y/L_x}{q\Omega}.
\end{equation}

\subsection{Wave equations}
We developed equations governing the excitation of SD waves in the inviscid
limit in paper~I. These follow directly from equations (\ref{eq:continuity})
and (\ref{eq:momentum}). As we will demonstrate, SD waves in the unstratified
shearing box exhibit little dependence on the vertical coordinate $z$ so that
we may consider vertically averaged quantities (indicated by being contained
within angle brackets $\langle\cdot\rangle_z)$ for which the wave equations
read
\begin{subequations}\label{eq:wave-equations}
\begin{multline}\label{eq:waverho}
(\mathcal{D}^2 - c^2\nabla^2 + \kappa^2)\zaver{\rho}
+ 2q\Omega\partial_y\zaver{p_x} = \\
- 2\Omega\zaver{\zeta} - \partial_x\zaver{\mathfrak{N}_x}
- \partial_y\zaver{\mathfrak{N}_y},
\end{multline}
\begin{multline}\label{eq:wavepx}
(\mathcal{D}^2 - c^2\nabla^2 + \kappa^2)\zaver{p_x}
+ 2q\Omega c^2\partial_y\zaver{\rho} = \\
c^2\partial_y\zaver{\zeta}
+ 2\Omega\zaver{\mathfrak{N}_y} + \mathcal{D}\zaver{\mathfrak{N}_x},
\end{multline}
and
\begin{multline}\label{eq:wavepy}
(\mathcal{D}^2 - c^2\nabla^2 + \kappa^2)\zaver{p_y} = \\
- c^2\partial_x\zaver{\zeta}
+ (q-2)\Omega\zaver{\mathfrak{N}_x} + \mathcal{D}\zaver{\mathfrak{N}_y}.
\end{multline}
\end{subequations}
Here, $\kappa^2=2(2-q)\Omega^2$ is the square of the epicyclic frequency and
we have introduced the divergence of the nonlinear stress tensor
$\vec{\mathfrak{N}}=\nabla\cdot\vec{T}$ as a short-hand to denote the
nonlinear source terms.

In paper~I we postulated that as far as wave excitation is concerned the
dominant source term in each of the three wave equations is the linear one
appearing first on the right hand sides of (\ref{eq:wave-equations}) and
involving the quantity
\begin{equation}\label{eq:zeta}
\zaver{\zeta} = \partial_x\zaver{p_y} - \partial_y\zaver{p_x}
+ (q-2)\Omega\zaver{\rho}
\end{equation}
which we refer to as pseudo potential vorticity (or PPV for short).
An important property of PPV is that apart from advection by the linear shear,
it varies in time due to nonlinear stresses only,
\begin{equation}\label{eq:dzeta}
\mathcal{D}\zaver{\zeta} = \partial_x\zaver{\mathfrak{N}_y}
- \partial_y\zaver{\mathfrak{N}_x},
\end{equation}
and is thus conserved for a linear SD wave.
Note also that in this limit the change in PPV is equal to the
change of potential vorticity, the difference being at most second order
in the wave amplitude.

\subsection{Decomposition into shearing waves}
In the periodic shearing sheet a suitable Fourier basis is given by plane
waves with, in the coordinate system adopted, time dependent radial wave
numbers. The expansion of any (vertically averaged) fluid variable $f$ in this
basis reads
\begin{equation}\label{eq:fourier-basis}
\zaver{f}(x,y,t) = \sum_{n_x,n_y}\!
\hat{f}(t)\,\exp\Big[\rmi k_x(t)x + \rmi k_y y\Big],
\end{equation}
where $\hat{f}$ is a complex valued wave amplitude and where the wave numbers
\begin{displaymath}
k_x(t) = \left(\frac{2\pi n_x}{L_x}\right)
+ q\Omega t\left(\frac{2\pi n_y}{L_y}\right)
\quad\mathrm{and}\quad
k_y = \frac{2\pi n_y}{L_y}.
\end{displaymath}
Each of these so-called shearing waves \citep{thomson_stability_1887} is
uniquely labelled by the pair of integral numbers
\mbox{$n_x,n_y\in\mathbb{Z}$}.
The radial wave numbers depend linearly on time such that
\begin{displaymath}
\der{k_x}{t} = q\Omega k_y,
\end{displaymath}
which is a consequence of advection by the linear background shear. 

Because \mbox{$k_x(t)/k_y$} increases monotonically if \mbox{$q\Omega>0$},
every shearing wave evolves from being leading, i.e.\ \mbox{$k_x(t)/k_y<0$},
to being trailing, i.e.\ \mbox{$k_x(t)/k_y>0$}, as time progresses from
\mbox{$t=-\infty$} to \mbox{$t=\infty$}. The change from leading to trailing
is referred to as `swing' and occurs at
\begin{equation}\label{eq:little-shearing-time}
t_\mathrm{s} = -\frac{n_x}{n_y}\delta T_\mathrm{s}
\end{equation}
From linear theory we expect wave excitation to occur at the time of the
swing. From (\ref{eq:little-shearing-time}) we see that different shearing
waves swing from leading to trailing at different times so that the excitation
process consists of a series of swings that are separated, for a given $k_y$,
by the fixed time interval
\mbox{$\delta{}t_\mathrm{s}=\delta{}T_\mathrm{s}/n_y$}.

Expanded in the Fourier basis (\ref{eq:fourier-basis}), the SD wave equations
(\ref{eq:wave-equations}) for a single shearing wave read
\begin{subequations}\label{eq:fourier-wave-equations}
\begin{multline}\label{eq:fourier-wave-rho}
\dder{\hat{\rho}}{t} + \left[\vec{k}^2(t)c^2 + \kappa^2\right]\hat{\rho}
+ 2q\Omega\rmi k_y\hat{p}_x = \\
- 2\Omega\hat{\zeta} - \rmi k_x(t)\hat{\mathfrak{N}}_x
- \rmi k_y\hat{\mathfrak{N}}_y,
\end{multline}
\begin{multline}\label{eq:fourier-wave-px}
\dder{\hat{p}_x}{t} + \left[\vec{k}^2(t)c^2 + \kappa^2\right]\hat{p}_x
+ 2q\Omega c^2\rmi k_y\hat{\rho} = \\
c^2\rmi k_y\hat{\zeta} + 2\Omega\hat{\mathfrak{N}}_y
+ \der{\hat{\mathfrak{N}}_x}{t},
\end{multline}
and
\begin{multline}\label{eq:fourier-wave-py}
\dder{\hat{p}_y}{t} + \left[\vec{k}^2(t)c^2 + \kappa^2\right]\hat{p}_y = \\
- c^2\rmi k_x\hat{\zeta} + (q-2)\Omega\hat{\mathfrak{N}}_x
+ \der{\hat{\mathfrak{N}}_y}{t}.
\end{multline}
\end{subequations}

\section{Simulations}\label{sec:simulations}
The steady state solution for the non-stratified shearing box, for which
the density is uniform and the velocity takes the from of a linear shear,
is subject to the magneto-rotational instability
when a relatively weak magnetic field with zero net flux is introduced. This
ultimately results in the production of turbulence that is powered by the
linear shear and attains a statistically steady state
\citep{1995ApJ...440..742H,1995ApJ...446..741B,1996ApJ...463..656S}.

We have carried out local, three-dimensional simulations of accretion disk
turbulence induced by the MRI. The purpose of these simulations is threefold.
Firstly to study and quantify the mode of excitation of SD waves in a
turbulent shearing box, secondly to justify some of the simplifying
assumptions that were made in order to proceed with the theoretical analysis
of the excitation mechanism given in paper~I, and thirdly to provide results
which can be tested against predictions made from this theory (see Section
\ref{sec:comparison}) so that the nature of the phenomenon and the implied
dependence on physical parameters can be confirmed.

\subsection{Setup}
We use the Pencil Code \citep[see e.g.][]{Brandenburg2002471} to solve the MHD
equations in the shearing box. The numerical method relies on using
sixth-order central finite differences for the evaluation of spatial
derivatives and a third-order Runge-Kutta scheme for time integration. The
method is stabilized by imposing explicit diffusion coefficients in all
evolution equations. The shearing box boundary conditions are implemented via
sixth-order polynomial interpolation. The code solves the isothermal MHD
equations in the shearing box approximation
(\ref{eq:continuity}--\ref{eq:induction}) in terms of mass density $\rho$,
fluid velocity $\vec{u}$, and, to ensure that $\nabla\cdot\vec{B}=0$, the
induction equation is rewritten in terms of the
magnetic vector potential $\vec{A}$ with $\vec{B}=\nabla\times\vec{A}$.
Details are given in Appendix~\ref{app:pencil-code}.

The setup is similar to the one used for a code comparison in
\cite{2007A&A...476.1123F}. We choose a box size of $(L_x,L_y,L_z)=(H,4H,H)$
and a resolution of $(N_x,N_y,N_z)=(128,512,128)$. \cite{2007A&A...476.1123F}
showed that in the zero-net-flux case, the level of MRI turbulence critically
depends on the magnetic Prandtl number $\mathrm{Pm}=\nu/\eta$, i.e.\ the ratio
of kinematic viscosity to magnetic resistivity. Here we set
$\nu=32\cdot{}10^{-5}H^2\Omega$ and $\eta=8\cdot{}10^{-5}H^2\Omega$ such that
$\mathrm{Pm}=4$. In this case we observe MRI turbulence that is, as in
\cite{2007A&A...476.1123F}, sustained for several hundred orbits and expected
to be numerically resolved for the values of the diffusion coefficients that
we impose.

\subsection{Volume averages}

\begin{figure}
\begin{center}
\includegraphics[width=\columnwidth]{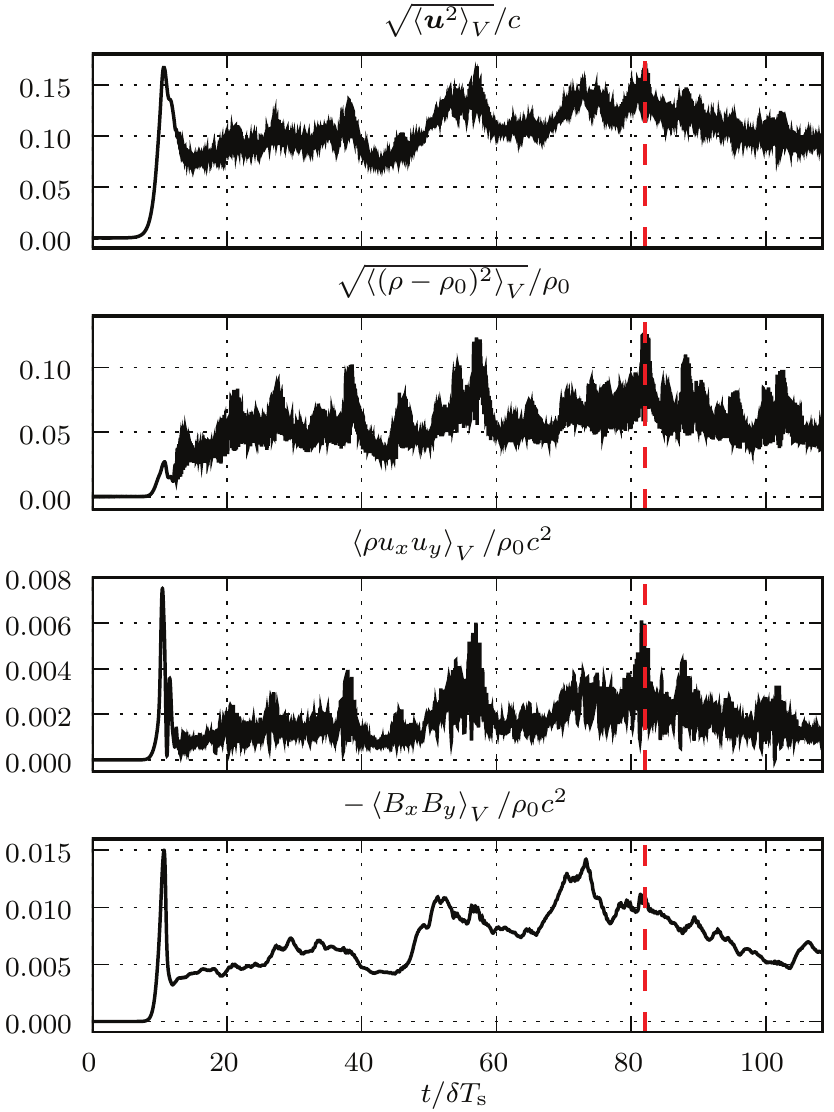}
\end{center}
\caption{Temporal evolution of root mean squared velocity and density
fluctuations (upper two panels) and of the volume averaged Reynolds and
Maxwell stress (lower two panels). The red dashed line corresponds to
$t=82\,\delta T_\mathrm{s}$ at which point the largest density fluctuation
occurs.}
\label{fig:time-series}
\end{figure}

In Fig.~\ref{fig:time-series} we plot the root mean squared velocity and
density fluctuations, calculated as volume averages over the box, and  volume
averages of the Reynolds and Maxwell stress as functions of time. The initial
peak that as seen in the velocity fluctuations and in the Reynolds and Maxwell
stresses is due to exponential growth of axisymmetric MRI modes. After roughly
4 Orbits, this so-called channel solution
is torn apart by secondary, non-axisymmetric instabilities
(see e.g.\ \citealt{2009MNRAS.tmp..192L} and references therein)
and the system enters a quasi steady state of sustained MRI turbulence.

The velocity fluctuations in this state are an order of magnitude less in
amplitude than the speed of sound. In spite of the strongly subsonic nature of
the turbulence we observe density fluctuations with a temporal mean of about 6
per cent and peak values of more than 15 per cent of the background value. We
will see that the peak density fluctuations are due to high-frequency,
non-axisymmetric SD waves that are excited via the mechanism
discussed in paper~I. Because these waves are trailing, they lead to outward
angular momentum transport. This manifests itself in a strong correlation
between the density fluctuations and the mean Reynolds stress. At \emph{peak}
values, the Reynolds stress may even become comparable to the Maxwell stress
although we note that the former fluctuates rapidly in time so that the net
angular momentum transport due to SD waves is less significant
than it may appear at first glance. We will attempt to quantify the amount of
angular momentum transport due to SD waves later in this paper.

We note that the Maxwell stress fluctuates in time much less rapidly than, and
is only very weakly correlated with, the other (purely hydrodynamic)
quantities. This is because SD waves are non-magnetic and the
Maxwell stress varies on a time scale comparable to the turn over time scale
of the turbulence which in this case is much longer than the inverse
SD wave frequency.

\subsection{Flow structure in real space}\label{sec:sim-real}

\begin{figure}
\begin{center}
\includegraphics[width=\columnwidth]{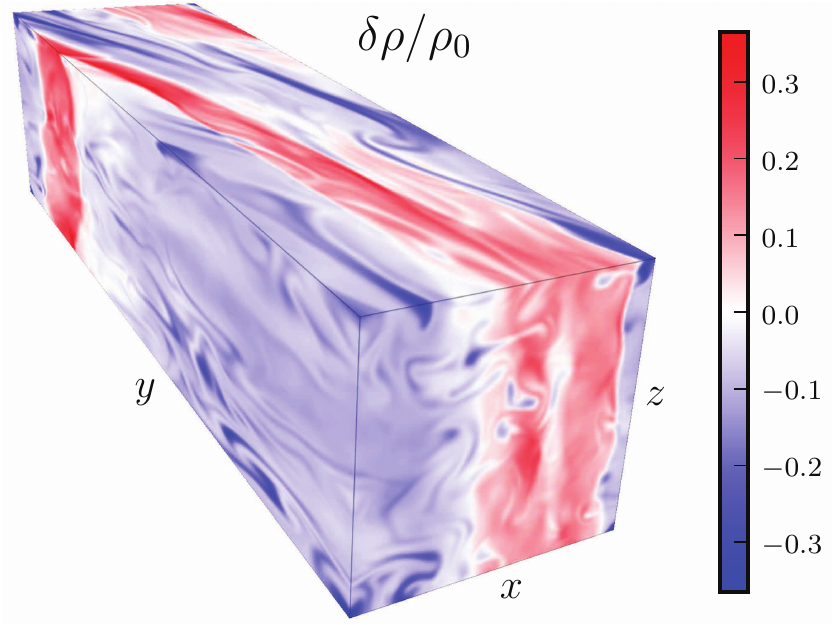}
\end{center}
\caption{Pseudo-colour image of mass density on the faces of the computational
domain. This snapshot is taken at \mbox{$t=82\,\delta{}T_\mathrm{s}$} where we
observe the largest peak in the volume averaged density fluctuations shown in
Fig.~\ref{fig:time-series}.}
\label{fig:cube}
\end{figure}

\begin{figure*}
\begin{center}
\includegraphics[width=\textwidth]{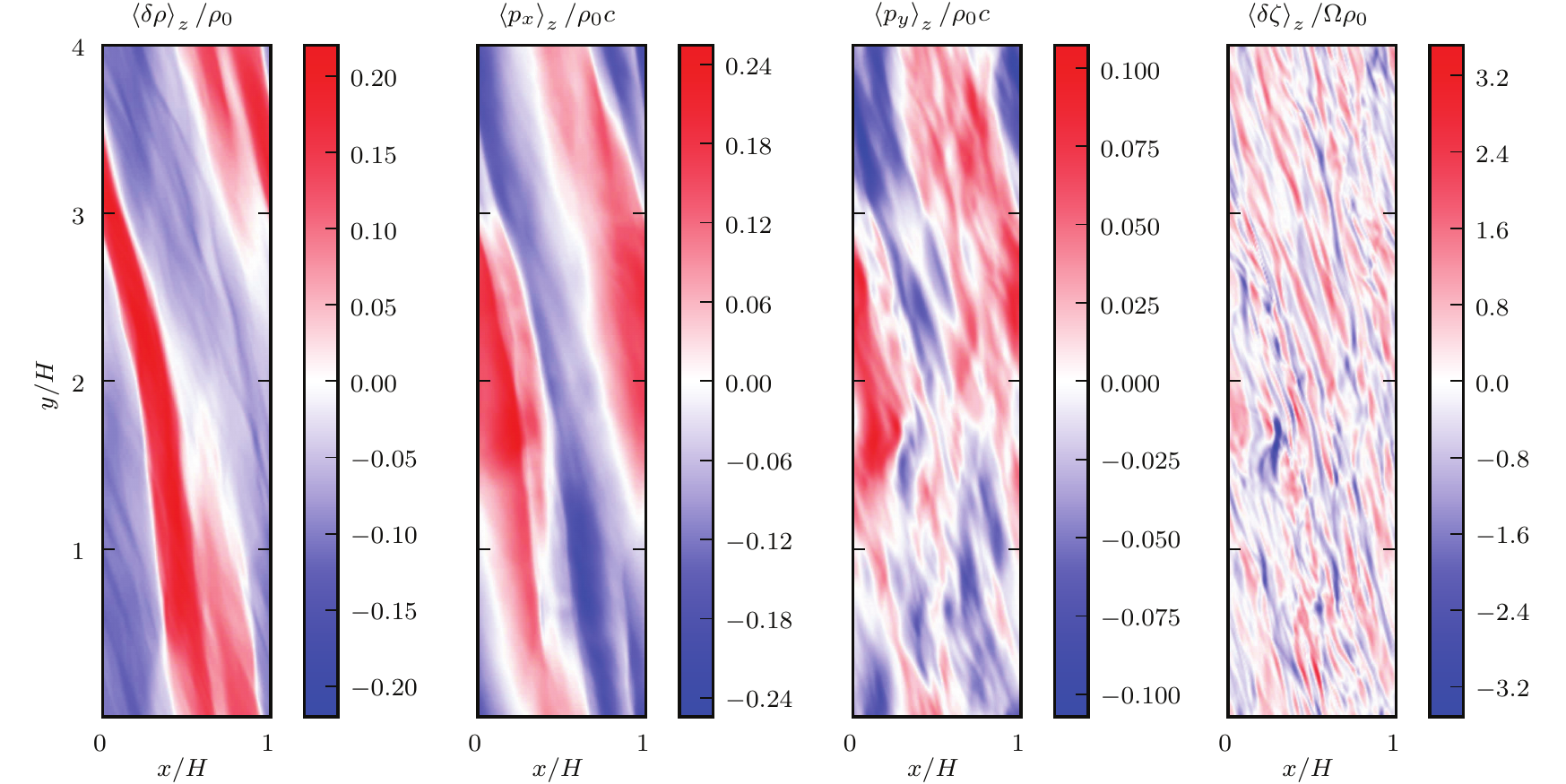}
\end{center}
\caption{Pseudo colour images of $z$-averaged quantities at the same instance
in time as in Fig.~\ref{fig:cube}. Non-axisymmetric waves are clearly visible
in \mbox{$\zaver{\rho}$} and \mbox{$\zaver{p_x}$}, and to a lesser extent in
\mbox{$\zaver{p_y}$}. There will be a system of regularly repeating wave
crests when boxes are stacked together to fill space by application of the
periodicity conditions. In contrast to this, the waves are not visible in
\mbox{$\zaver{\delta\zeta}$}, where \mbox{$\delta\zeta=\zeta-\zeta_0$} is the
PPV deviation from the constant background value
\mbox{$\zeta_0=(q-2)\Omega\rho_0$}.}
\label{fig:zaverage}
\end{figure*}

The typical large scale three-dimensional structure of the turbulent flow is
depicted in Fig.~\ref{fig:cube} where we show the mass density field projected
onto the faces of the computational domain\footnote{An animated version of
Fig.~\ref{fig:cube} and Fig.~\ref{fig:zaverage} below is provided as part of
the on-line supplements to this article}. This snapshot is taken at
\mbox{$t=82\,\delta{}T_\mathrm{s}$} where we observe the largest peak in the
volume averaged density fluctuations shown in Fig.~\ref{fig:time-series}.

The wave crests are seen to propagate at an inclined angle within the
horizontal $xy$-plane and display little dependence on $z$. In order to ease
the subsequent analysis, as in paper~I, we may thus integrate over $z$ and
will look at vertically averaged quantities only from now on.

In Fig.~\ref{fig:zaverage} we plot false colour images of such vertical
averages at the same instance in time as shown in Fig.~\ref{fig:cube}. Large
scale wave crests are clearly visible in $\zaver{\rho}$, $\zaver{p_x}$ and, to
a lesser extent, in $\zaver{p_y}$. The waves are trailing and always occur in
pairs where one wave propagates in a direction opposite to the other wave.
That this has to be the case is clear from symmetry considerations of the
shearing sheet and we recall that our linear theory predicts pairwise
excitation of waves with equal amplitude but oppositely directed wave vectors.

In contrast to $\zaver{\rho}$, $\zaver{p_x}$,
and $\zaver{p_y}$, the pseudo potential vorticity
\begin{equation*}
\zaver{\zeta} = \partial_x\zaver{p_y} - \partial_y\zaver{p_x}
+ (q-2)\Omega\zaver{\rho}
\end{equation*}
does not display a corresponding large
scale wave structure but instead consists of smaller scale turbulent
structures that are, as is characteristic for accretion disc turbulence,
elongated in the shearwise direction. The absence of large scale wave
structure in $\zaver{\zeta}$ is reminiscent of the fact that $\zaver{\zeta}$
is conserved for a linear wave, see (\ref{eq:dzeta}), and the justification
for why we are allowed to treat $\zaver{\zeta}$ as a source term for wave
excitation, as was done in paper~I, in the first place.

\subsection{Evolution of single shearing wave}\label{sec:evo-single-wave}

\begin{figure}
\begin{center}
\includegraphics[width=\columnwidth]{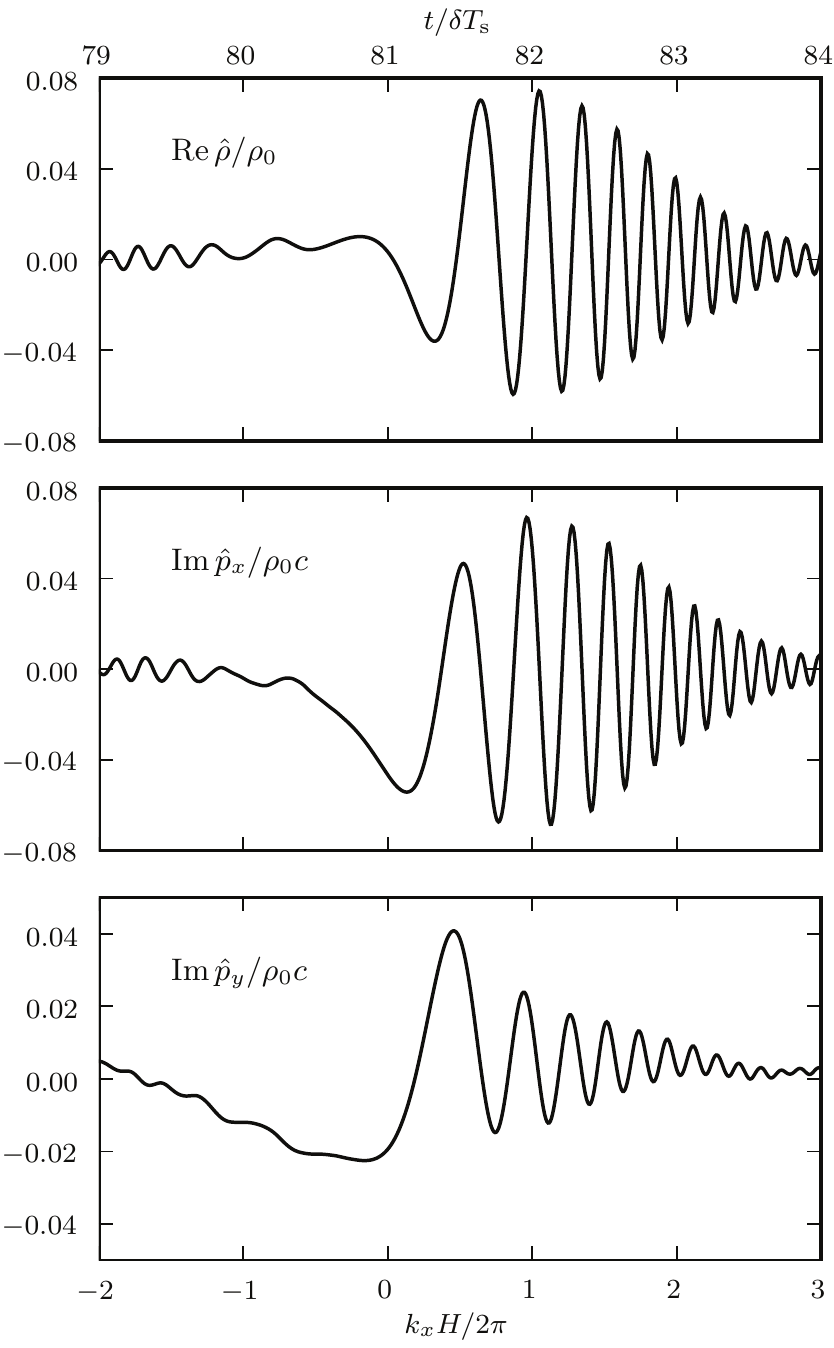}
\end{center}
\caption{Fourier amplitudes of density and linear momenta for a single
shearing wave with azimuthal wave number $k_y=2\pi/L_y$ as it swings from
leading to trailing. Excitation occurs when $k_x(t)=0$ or, equivalently, at
\mbox{$t=81\,\delta{}T_\mathrm{s}$}. This is thus the same wave that is shown
in Fig.~\ref{fig:cube} and Fig.~\ref{fig:zaverage}.}
\label{fig:single-wave}
\end{figure}

A more revealing picture of the evolution of SD waves as
observed during the simulation may be obtained in Fourier space. For this
purpose we will look at the discrete Fourier transform (DFT) of vertically
averaged quantities as shown in Fig.~\ref{fig:zaverage}. Taking a
two-dimensional spatial DFT in the shearing sheet is non-trivial due to the
presence of the linear background shear and requires a coordinate
transformation to shearing coordinates, see Appendix~\ref{app:fourier}.
This coordinate transformation can be done
very conveniently by first taking the DFT along the periodic $y$-direction,
then applying an $x$-dependent phase shift to achieve full periodicity in $x$,
and finally taking the DFT along the $x$-direction. Further details are given
in the Appendix~\ref{app:fourier}. This procedure
allows us to uniquely assign a time-dependent radial wave number $k_x(t)$ as
well as a (constant) azimuthal wave number $k_y$ to a given Fourier amplitude
obtained from the DFT, enabling us to follow individual plane waves as they
swing from leading to trailing during the course of the simulation.

In Fig.~\ref{fig:single-wave} we plot the evolution of the wave amplitudes of
the $\hat{\rho}$, $\hat{p}_x$, and $\hat{p}_y$ for a
single shearing wave with azimuthal wave number $k_y=2\pi/L_y$. Excitation
occurs when the wave swings from leading to trailing, i.e. when
$k_x(t)\approx{}0$, which for this particular wave happens at
\mbox{$t=81\,\delta{}T_\mathrm{s}$}. The maximum density amplitude of
$\mathrm{Re}\,\hat{\rho}\approx{}0.08\,\rho_0$ is obtained when
$k_x(t)\approx{}2\pi/H$ or, equivalently, around
\mbox{$t=82\,\delta{}T_\mathrm{s}$}, i.e.  roughly one shearing time later. In
Fig.~\ref{fig:single-wave} we are thus looking at the Fourier representation
of the wave seen in Fig.~\ref{fig:cube} and Fig.~\ref{fig:zaverage} and thus
at that particular wave responsible for the large peak in the volume
averaged density fluctuations indicated by the dashed red line in
Fig.~\ref{fig:time-series}.

We note that the wave undergoes significant damping for
\mbox{$k_x(t)H\gtrsim{}2\pi$}.
This decay is in stark contrast to linear theory which
either predicts an \emph{algebraic} decrease (for $\hat{p}_y$) or even
increase (for $\hat{\rho}$ and $\hat{p}_x$) of the wave amplitude with
time (see paper~I) and therefore needs to be attributed to nonlinear effects.
We will comment on this further in Section~\ref{sec:comparison}.

\begin{figure}
\begin{center}
\includegraphics[width=\columnwidth]{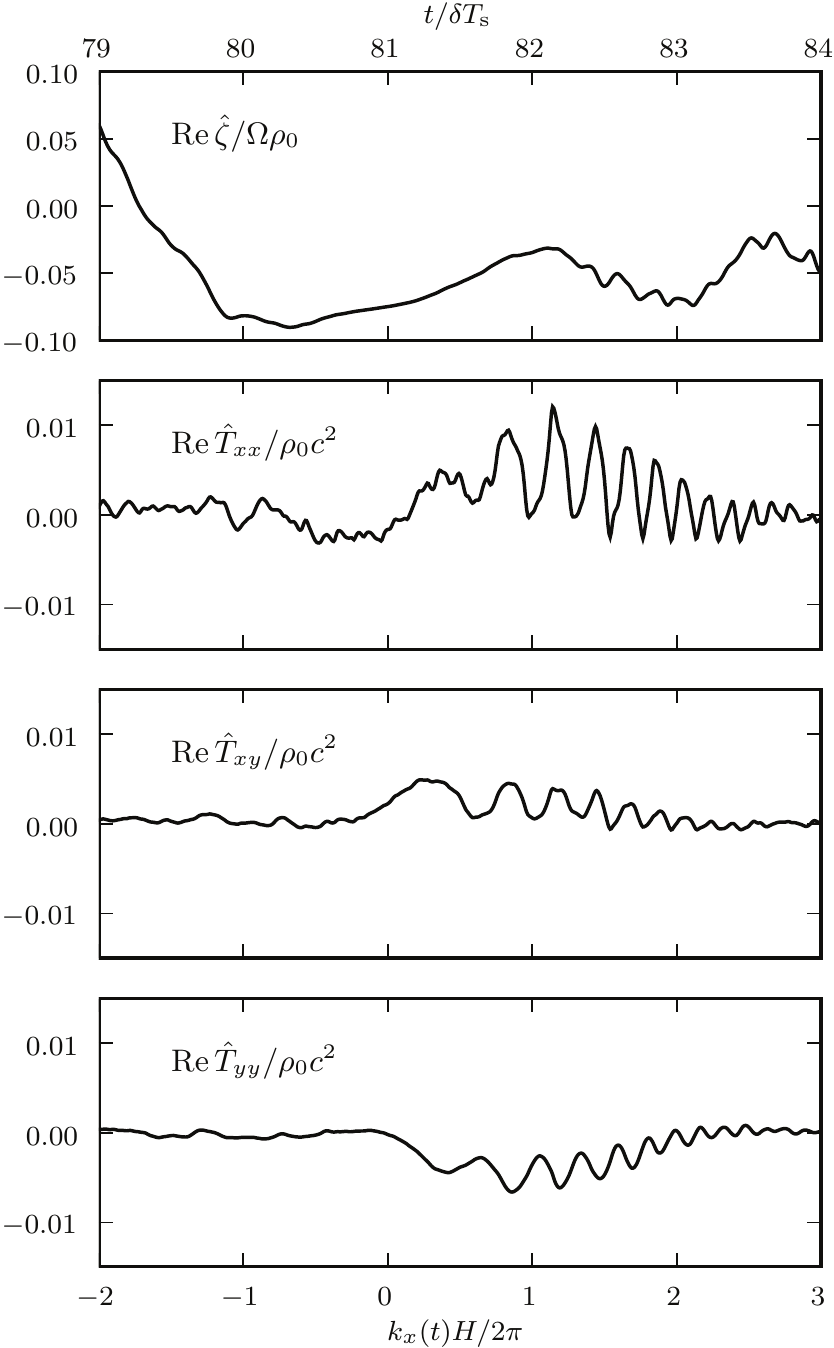}
\end{center}
\caption{Fourier amplitudes of the pseudo potential vorticity and the
components of the nonlinear stress tensor. All of these appear as sources on
the right hand side of the SD wave-equations
(\ref{eq:wave-equations}).}
\label{fig:single-wave-sources}
\end{figure}

The simulation allows us not only to monitor the wave amplitudes $\hat{\rho}$,
$\hat{p}_x$, and $\hat{p}_y$ for a single shearing wave as it evolves in time,
but also the various source terms occurring on the right sides of the
SD wave equations (\ref{eq:fourier-wave-equations}), and to assess the
relative importance of these terms for wave excitation.

There are both linear source terms involving the pseudo potential vorticity as
well as nonlinear terms involving the vector
$\vec{\hat{\mathfrak{N}}}=\rmi\vec{k}\cdot\vec{\hat{T}}$, where
$\hat{T}_{ij}$ is the Fourier transform of the nonlinear stress tensor
(\ref{eq:stress-tensor}).

In Fig.~\ref{fig:single-wave-sources} we plot the
Fourier amplitudes of $\hat{\zeta}$ and the $\hat{T}_{ij}$ for the wave shown
in Fig.~\ref{fig:single-wave}. We see that $\hat{\zeta}$ is -- compared to the
$\hat{\rho}$, $\hat{p}_x$, $\hat{p}_y$, and the
nonlinear stresses $\hat{T}_{ij}$ -- a slowly varying function of time. When
wave excitation occurs around $k_x(t)=0$ there is no sign of oscillating
behaviour in $\hat{\zeta}$. This is a manifestation of PPV conservation
to linear order, see (\ref{eq:dzeta}).

Furthermore, at the time of wave excitation, the amplitude of $\hat{\zeta}$ is
more than an order of magnitude larger than the amplitude of any of the
components of the nonlinear stress tensor $\hat{T}_{ij}$ suggesting, as these
quantities measure their relative contributions to the source terms in
equations (\ref{eq:fourier-wave-equations}), that it is indeed the pseudo
potential vorticity, and thus the linear source term, that is primarily
responsible for wave excitation. We will provide more evidence for this claim
in Section~\ref{sec:comparison}.

\section{Comparisons with asymptotic theory}\label{sec:comparison}

The evolution of SD waves in the shearing sheet is described
mathematically by the time-dependent forced oscillator equations
(\ref{eq:fourier-wave-equations}). Wave excitation is due to the presence of
inhomogeneous source terms in these equations and is found -- as demonstrated
in the preceding section -- to occur at the swing stage, i.e. when the time
dependent radial wave number $k_x(t)=0$.

In paper~I we derived analytic solutions to the governing
equations (\ref{eq:fourier-wave-equations}) that are able to capture this
swing excitation. These solutions are asymptotic in the parameter
\begin{equation*}
\epsilon = \frac{q\Omega k_y c}{k_y^2 c^2 + \kappa^2},
\end{equation*}
which is small in both the low and the high azimuthal wave number limit, given
by $k_yc\ll\kappa$ and $k_yc\gg\kappa$, respectively.  The asymptotic
solutions can be written as the sum of a non-oscillatory, vortical solution --
called the balanced solution -- to the inhomogeneous problem, and standard
WKBJ solutions to the free wave equations that contain the oscillatory part of
the solutions and describe the excited SD waves.

A key hypothesis in our analysis in paper~I was that the dominant source term
for wave excitation is the linear term involving the pseudo potential
vorticity $\hat{\zeta}$ and that nonlinear stresses contribute to the
excitation only indirectly by generating $\hat{\zeta}$ prior to the swing via
(\ref{eq:dzeta}). This assumption renders the analysis as essentially linear.
We have shown in Section~\ref{sec:evo-single-wave} that the pseudo potential
vorticity for a large amplitude SD wave observed in the
simulation is more than an order of magnitude greater in size than the
nonlinear stresses when the excitation occurs. This suggests that our
quasi-linear description of the excitation process is indeed valid and we are
now in a position to verify this assertion by comparing the observed wave
amplitudes with our predictions from linear theory.

Expressed in dimensional form, the leading order balanced solutions for
$\hat{\rho}$, $\hat{p}_x$, and $\hat{p}_y$ can be obtained from the relations
given in paper~I as
\begin{subequations}\label{eq:background-solutions}
\begin{align}
\label{eq:background-solutions-rho}
\bar{\rho} &= \hat{\zeta}\,\mathrm{Re}\left(
\frac{\rmi k_y c - 2\Omega}{\vec{k}^2 c^2 + \kappa^2 + 2q\Omega\rmi k_y c}
\right), \\
\label{eq:background-solutions-px}
\bar{p}_x &= \rmi\hat{\zeta}c\,\mathrm{Im}\left(
\frac{\rmi k_y c - 2\Omega}{\vec{k}^2 c^2 + \kappa^2 + 2q\Omega\rmi k_y c}
\right), \\
\label{eq:background-solutions-py}
\bar{p}_y &= \rmi\hat{\zeta}c\,\mathrm{Re}\left(
\frac{\rmi k_x c}{\vec{k}^2 c^2 + \kappa^2}
\right),
\end{align}
\end{subequations}
where the SD wave frequencies
\begin{equation*}
\omega = \sqrt{\vec{k}^2 c^2 + \kappa^2}
\quad\mathrm{and}\quad
\omega_{+} = \sqrt{\omega^2 + 2q\Omega\rmi k_y c}.
\end{equation*}
Note that these solutions include the full time dependence of the source.

We are, however, mainly interested in the oscillatory wave part of the
solutions, i.e.\ the WKBJ solutions to the homogeneous problem given in
paper~I, which may be written as
\begin{subequations}\label{eq:wave-solutions}
\begin{equation}
\label{eq:wave-solutions-rho}
\tilde{\rho} = -\zeta_\mathrm{s}\Omega^{-1}
\mathrm{Im}\left[\mathcal{A}_{+}\sqrt{\Omega/\omega_{+}}
\sin\left(\int_0^t\!\omega_{+}\,\rmd t'\right)\right]
\end{equation}
and
\begin{equation}
\label{eq:wave-solutions-px}
\tilde{p}_x = \rmi\zeta_\mathrm{s}H\,
\mathrm{Re}\left[\mathcal{A}_{+}\sqrt{\Omega/\omega_{+}}
\sin\left(\int_0^t\!\omega_{+}\,\rmd t'\right)\right]
\end{equation}
\end{subequations}
where
\begin{multline}\label{eq:wave-amplitude}
\mathcal{A}_{+} =
\frac{\Omega^{1/2}}{(k_y^2 c^2 + \kappa^2 + 2q\Omega\rmi k_y c)^{1/4}} \\
\left(\frac{2\Omega - \rmi k_y c}{\sqrt{k_y^2 c^2 + \kappa^2}}\right)
\sqrt{\frac{2\pi}{\epsilon}}\,\rme^{-\pi/4\epsilon}.
\end{multline}
The free WKBJ solution for $\hat{p}_y$ follows from PPV conservation,
\begin{displaymath}
\tilde{p}_y = \frac{\rmi k_y\tilde{p}_x + 2(2-q)\Omega\tilde{\rho}}{\rmi k_x}.
\end{displaymath}
In the above expressions, only the value of the pseudo potential vorticity at
the time of the swing (denoted by $\hat{\zeta}_\mathrm{s}$) enters, to which
the amplitude of the excited wave is directly proportional. This
amplitude is exponentially small in the asymptotic limit $\epsilon\ll{}1$.
However, we demonstrated in paper~I that our leading order asymptotic
solutions are remarkably accurate even for $\epsilon\lesssim{}1$, and in the
case of Keplerian shear effectively hold for the entire range of azimuthal
wave lengths because $\epsilon\le{}3/4$ for $0\le{}k_yH<\infty$.

\subsection{Comparison of wave amplitudes}
We are now in a position to test the linear theory of the excitation process
by comparing the asymptotic solutions (\ref{eq:background-solutions}) and
(\ref{eq:wave-solutions}) with the evolution of an individual shearing wave
observed in the simulation as shown in Fig.~\ref{fig:single-wave} and
Fig.~\ref{fig:single-wave-sources}. For this particular wave, the azimuthal
wave number $k_y=2\pi/L_y$ and thus $\epsilon\approx{}0.68$.

\begin{figure}
\begin{center}
\includegraphics[width=\columnwidth]{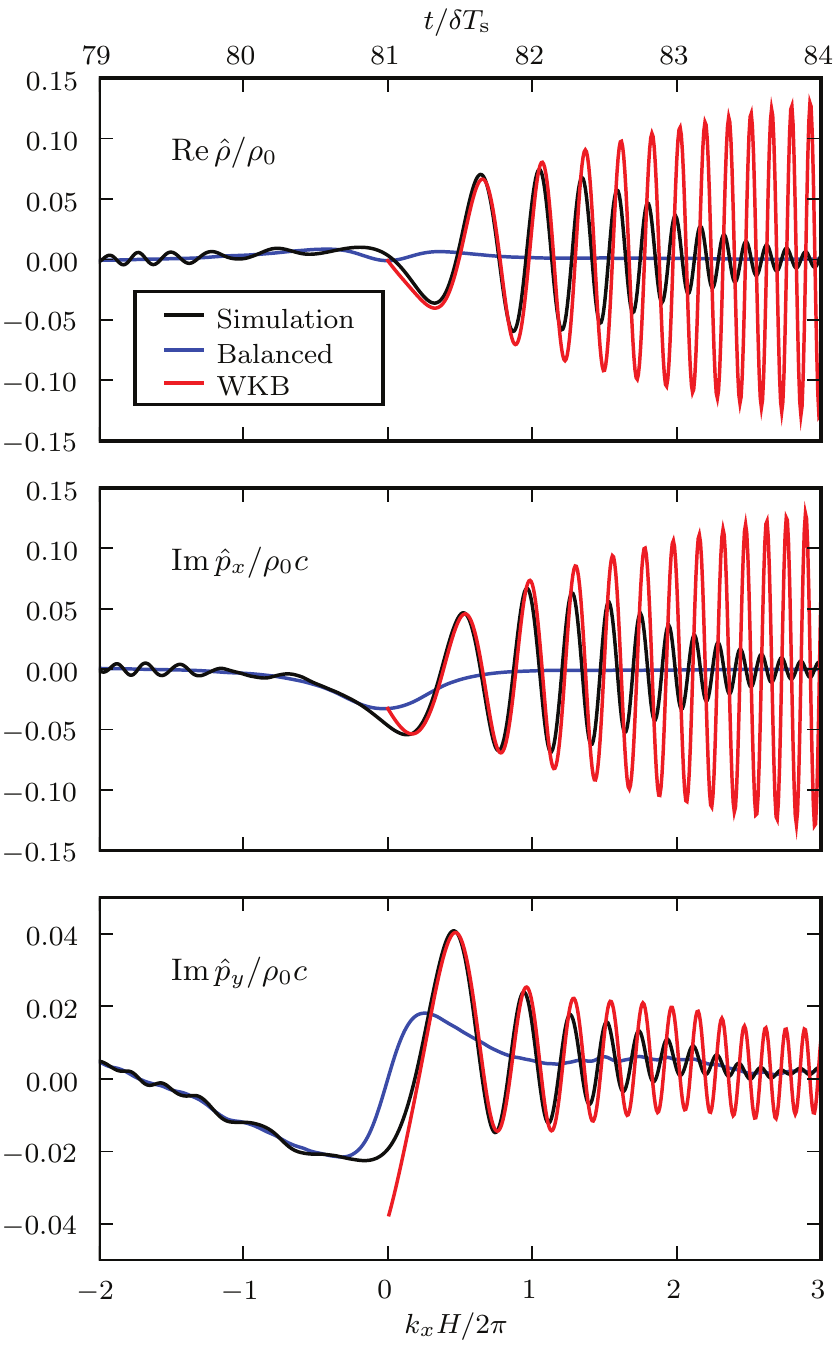}
\end{center}
\caption{Comparison between simulation and linear theory for the same shearing
wave as in Fig.~\ref{fig:single-wave}. The curves shown in each panel are the
wave amplitude as recorded during the simulation (black), the balanced
solution (blue), and the full WKBJ solution (red). For each variable, the WKBJ
amplitude in the upper panel is taken as is from linear theory and therefore
grows/decays algebraically with time.}
\label{fig:mode-history}
\end{figure}

The result is shown in Fig.~\ref{fig:mode-history}. As in the linear problem
discussed in paper~I, the numerical solution in the leading phase follows the
balanced solution very closely. In the trailing phase the WKBJ solution
provides an excellent match to the numerical solution for small radial wave
numbers $k_x\lesssim{}2\pi/H$. Roughly one shearing time after excitation,
when the radial wave length is equal to the scale height $H$, the numerical
solution starts to undergo significant damping whereas the amplitude of the
WKBJ solution continues to increase algebraically.  Empirically we find that
the decay of the numerical solution cannot simply be attributed to viscous
effects because the diffusion coefficients imposed for this simulation are at
least an order of magnitude too small.

We are thus lead to believe that SD waves are damped due to nonlinear
effects. It is clear that a linear shearing wave whose radial wave length
decreases as its amplitude increases eventually has to break and form a shock.
This phenomena has been already discussed in the context of disk physics
applied to e.g.\ Saturn's rings \citep{1978Icar...34..240G} and to spiral
density wakes of protoplanets \citep{2001ApJ...552..793G}. If damping is
mostly due to this effect, we conclude from Fig.~\ref{fig:mode-history} that
shock formation must happen already when the radial wave length is roughly
equal to $H$. This fits with the observation that SD waves usually appear as
weak shocks in the simulation, see Fig.~\ref{fig:cube} and
Fig.~\ref{fig:zaverage} as well as e.g.\ \citet{2005AIPC..784..475G}.

\subsection{The importance of PPV}

\begin{figure}
\begin{center}
\includegraphics[width=\columnwidth]{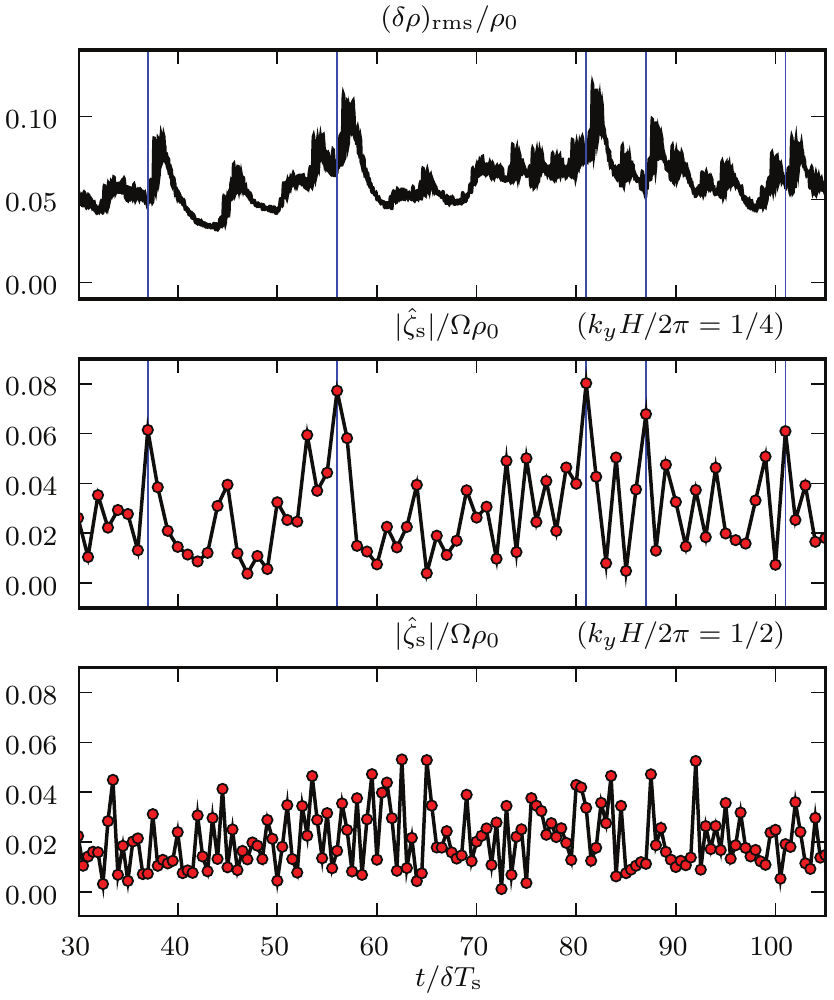}
\end{center}
\caption{Upper panel: Volume averaged density fluctuations as a function of
time. Here we have used high order binomial smoothing to get rid of the fast
oscillations seen in Fig.~\ref{fig:time-series}. Lower panel: Modulus of the
potential vorticity amplitude for the two smallest non-zero azimuthal wave
number in the system (\mbox{$k_y=2\pi/L_y$} and \mbox{$k_y=\pi/L_y$}). Each
red dot corresponds to an instant in time when there is a shearing wave with
\mbox{$k_x(t)=0$}. This happens once (twice) per shearing time for waves with
\mbox{$k_y=2\pi/L_y$} (\mbox{$k_y=\pi/L_y$}).}
\label{fig:zeta-ts}
\end{figure}

The results from the previous subsection indicate that linear theory is able
to account for the initial evolution of an excited SD wave
observed in the simulation. From (\ref{eq:wave-solutions}) we see that the
only dynamic quantity that enters the linear SD wave amplitude
is the pseudo potential vorticity at the swing state,
$\hat{\zeta}_\mathrm{s}$. If the particular wave discussed in the previous
subsection was not just a lucky coincidence, and linear theory holds in
general as far as the initial evolution is concerned, we thus expect a
correlation between $\hat{\zeta}_\mathrm{s}$ and, for instance, the root mean
squared density fluctuations,
\begin{equation*}
(\delta\rho)_\mathrm{rms} = \sqrt{\left<(\rho-\rho_0)^2\right>_V},
\end{equation*}
shown in Fig.~\ref{fig:time-series}.

In Fig.~\ref{fig:zeta-ts} we plot $(\delta\rho)_\mathrm{rms}$ as well as
$|\hat{\zeta}_\mathrm{s}|$ for the two shortest, non-zero azimuthal wave
numbers available in our computational domain, denoted here by
\mbox{$k_y^\mathrm{I}$} and \mbox{$k_y^\mathrm{II}$}, as a function of time.
The temporal correlation between $(\delta\rho)_\mathrm{rms}$ and
$|\hat{\zeta}_\mathrm{s}|$ for \mbox{$k_y=k_y^\mathrm{I}$} is remarkably
strong indeed.  In particular the peak density fluctuations are seen to be
delayed by roughly one shearing time with respect to the corresponding peak in
$|\hat{\zeta}_\mathrm{s}|$. This is because it takes roughly one shearing time
for an SD wave to amplify to full strength after excitation, see
Fig.~\ref{fig:mode-history}. This hardly obvious correlation is perhaps the
strongest evidence for the validity of our linear theory and the neglect of
nonlinear source terms in the calculation of the wave amplitude.

In contrast to the above, the temporal correlation between
$(\delta\rho)_\mathrm{rms}$ and $|\hat{\zeta}_\mathrm{s}|$ for
\mbox{$k_y=k_y^\mathrm{II}$} is much weaker even though the average magnitude
of $\hat{\zeta}_\mathrm{s}$ is still roughly 70 per cent of the average
magnitude for \mbox{$k_y=k_y^\mathrm{I}$}. On one hand, this is likely to be
the case because the correlation for \mbox{$k_y=k_y^\mathrm{II}$} is obscured
and therefore harder to see, but more importantly because swing excitation --
as we know from linear theory -- is simply a weaker effect for
\mbox{$k_y=k_y^\mathrm{II}$} than for \mbox{$k_y=k_y^\mathrm{I}$} due to the
exponential dependence of the WKBJ amplitudes on $k_y$, see
(\ref{eq:wave-amplitude}). We will comment on this further in the discussion
section.

\subsection{Angular momentum flux}

In paper~I we derived an expression for the radial angular momentum flux
associated with a single pair of linear SD waves propagating in
opposite directions,
\begin{equation}\label{eq:mom-flux-xi}
\left<F_x\right>_{yz} =
- 2 k_y c^2 \mathrm{Im}\Big(\tilde{\xi}_x^\ast\tilde{\rho}\Big),
\end{equation}
see Goodman \& Ryu. The angular brackets denote an average over azimuth, and
as in paper~I we take $k_y\ge 0$. The radial Lagrangian displacement
$\hat{\xi}_x$ is defined via
\begin{displaymath}
\der{\hat{\xi}_x}{t} = \frac{\hat{p}_x}{\rho_0}.
\end{displaymath}
Note that (\ref{eq:mom-flux-xi}) only includes the oscillatory wave part of
the Fourier amplitudes, i.e.\ we do not account for the supposedly small
contributions to the total angular momentum transport resulting from the
slowly varying balanced solutions (\ref{eq:background-solutions}).

Since from a numerical point of view it is difficult to unambiguously separate
the oscillatory wave part of $\hat{\xi}_x$, it is -- for diagnostic purposes
-- more convenient to use the modified angular flux
\begin{equation}\label{eq:mom-flux-py}
\left<F'_x\right>_{yz} =
\frac{2 k_y c^2}{2(2-q)\Omega\rho_0}\mathrm{Im}(\tilde{p}_y^\ast\tilde{\rho})
\end{equation}
which does not involve the Lagrangian displacement. The two alternative
expressions for the angular momentum flux, (\ref{eq:mom-flux-xi}) and
(\ref{eq:mom-flux-py}), are found to agree when averaged in time over one
oscillation period.

It is now of interest to compare the angular momentum flux
(\ref{eq:mom-flux-py}) associated with a single pair of SD waves
observed in the simulation with linear theory by substituting the free wave
solutions
(\ref{eq:wave-solutions}) into
(\ref{eq:mom-flux-py}). The wave part of the Fourier amplitudes
obtained from the simulations is calculated by subtracting the
balanced solutions (\ref{eq:background-solutions}).

\begin{figure}
\begin{center}
\includegraphics[width=\columnwidth]{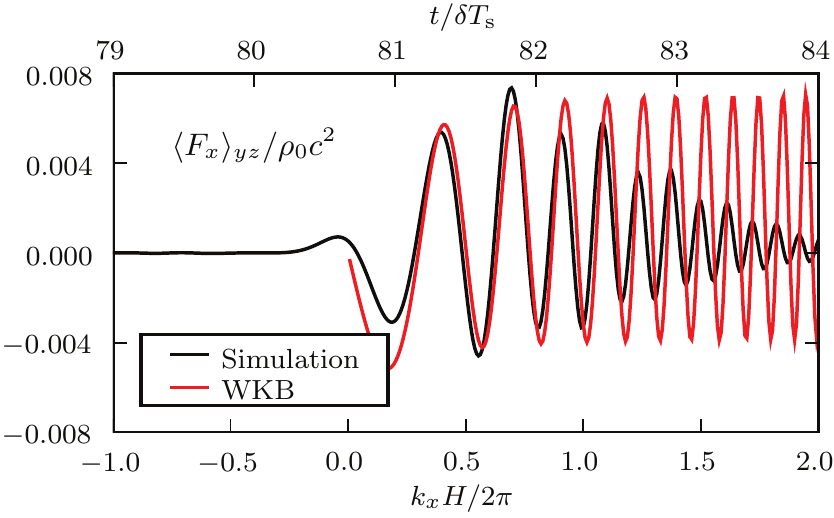}
\end{center}
\caption{Angular momentum flux associated with the SD wave
displayed in Fig.~\ref{fig:mode-history}.}
\label{fig:wave-action}
\end{figure}

The result is shown in in Fig.~\ref{fig:wave-action} where we plot the
modified angular momentum flux associated with the SD wave
displayed in Fig.~\ref{fig:mode-history}.
As expected there is essentially no angular momentum transport prior to
excitation, i.e.\ as long the waves are leading. In the trailing phase, after
swing excitation has occurred, the angular momentum flux oscillates rapidly
due to interference between the forward and the backward travelling wave.
However, it does so around a positive mean, which means that there is net
\emph{outward} angular momentum transport.

In contrast to linear theory, the angular momentum flux observed in the
simulation is strongly damped for moderately large wave numbers due to
nonlinear effects. Interestingly enough, however, before the onset of damping
the flux for a single pair of shearing waves observed in the simulation
reaches a maximum value that is very close to what is predicted by linear
theory for the far trailing regime.  In spite of nonlinear damping, there
will be a steady flux of angular momentum due to successive swings of shearing
waves and we can thus use linear theory to make an order of magnitude estimate
of the angular momentum transport due to SD waves.

In linear theory we calculated the net angular momentum flux in the far
trailing regime explicitly in paper~I. Averaged over one oscillation period,
this flux attains a constant non-zero value given by
\begin{equation}\label{eq:limiting-value-xi}
\overline{\langle F_x\rangle}_{yz} =
\frac{2\pi|\hat{\zeta}_\mathrm{s}|^2 H^2}{q\rho_0}
\frac{\Omega(k_y^2 c^2 + 4\Omega^2)\,\rme^{f(\epsilon)}\,\rme^{-\pi/2\epsilon}}
{\left[(k_y^2 c^2 + \kappa^2)^2 + (2q\Omega k_y c)^2\right]^{3/4}},
\end{equation}
as $k_x(t)\to\infty$. In the above equation,
\begin{displaymath}
f(\epsilon) = 1 - \frac{\tan^{-1}(2\epsilon)}{2\epsilon}.
\end{displaymath}

In order to estimate the amount of angular momentum transport due to
SD waves in the simulation, we thus require the swing PPV amplitude
$\hat{\zeta}_\mathrm{s}$.
For shearing waves with the longest azimuthal wave length available
in our computational domain, this information can be extracted from
Fig.~\ref{fig:zeta-ts} where we plot the modulus of the swing PPV amplitude
for all shearing waves that we
observe to swing from leading to trailing.
Making an ensemble average over all swings displayed in
Fig.~\ref{fig:zeta-ts} we obtain
\begin{displaymath}
|\hat{\zeta}_\mathrm{s}|^2 \approx
10^{-3}\Omega^2\rho_0^2.
\end{displaymath}
Substituting this into (\ref{eq:limiting-value-xi}) yields
\begin{displaymath}
\lim_{k_x/k_y\to\infty}\overline{\left<F_x\right>}_{yz} \approx
2.5\cdot 10^{-4}\rho_0 c^2.
\end{displaymath}
From this rough estimate we conclude that the mean angular momentum transport
due to SD waves in the present simulation is possibly less but still
significant compared to the turbulent Reynolds stress, see
Fig.~\ref{fig:time-series}.

\section{Summary and discussion}\label{sec:discussion}
In this paper we have presented fully three-dimensional simulations of
compressible MRI turbulence in which we observed the excitation of large
scale, non-axisymmetric SD waves. The root mean square density fluctuations
associated with these waves were shown to be correlated with the degree of
turbulent activity measured by the
\citet{1973A&A....24..337S} $\alpha$ parameter,
\begin{displaymath}
\alpha = \frac{\left<B_x B_y - \rho u_x u_y\right>_V}{\rho_0 c^2}.
\end{displaymath}

We performed a detailed analysis of the excitation of these waves by following
the time dependent evolution of the Fourier transforms of the participating
state variables. Their evolution during the excitation phase of the associated
SD wave was found to be in excellent agreement with the WKBJ theory developed
in paper~I. However, shortly after the attainment of the maximum wave
amplitude, the waves are rapidly damped. This damping is consistent with the
sharp density profiles observed in the simulations which are indicative
of the existence of shocks. Shocks are expected even under
conditions of weak excitation because as shearing waves propagate their radial
wavelength shortens, resulting in the onset of
nonlinearity and shock formation \citep[e.g.][]{2001ApJ...552..793G}. Thus
the waves are always likely to be observed in the nonlinear regime. Our
results indicate that for activity levels corresponding to
\mbox{$\alpha\sim{}5\times{}10^{-3}$}, being at the lower range of magnetic
activities normally considered, shock formation occurs
when their radial wavelength is comparable to the
scale height.

The excited waves are trailing and associated with an outward angular
momentum flux. The calculations performed here are consistent with the WKBJ
theory presented in paper~I in indicating that this outward flux scales in the
same way as the Reynolds stress. This is supported by the fact that the
density fluctuations associated with the waves are correlated with it. The
level of angular momentum transport associated with SD waves is
estimated to be a significant fraction of that
due to the turbulent Reynolds stress $\left<-\rho u_x u_y\right>_V$.

By consideration of the source terms for the wave excitation, we identified
those associated with the pseudo potential vorticity (PPV) as the dominant
ones as was assumed in paper~I. Wave like structure was not seen in the
PPV field, see Fig.~\ref{fig:zaverage},
indicating that
this could act consistently as an independent source generated by background
turbulent fluctuations.

For the simulation discussed in detail, SD wave excitation is found to be most
effective for waves with the smallest azimuthal wave number available in the
computational domain. This can be understood from the theory
presented in paper~I. There it was found that the amplitude of an excited
SD wave is directly proportional to the Fourier amplitude of the
pseudo potential vorticity at the swing stage, $\hat{\zeta}_\mathrm{s}$, while
also having an exponential dependence on the azimuthal wave number $k_y$
through the parameter
\begin{equation*}
\epsilon = \frac{q\Omega k_y c}{k_y^2 c^2 + \kappa^2}.
\end{equation*}
If the latter dependence alone is
important it implies that, for a Keplerian rotation law, SD wave production
will be most effective for $k_y\sim{}k_y^\mathrm{opt}=\Omega/c$. This is longer
than the longest possible azimuthal wavelength in the shearing box considered
here with $L_y=4H$, so that SD wave production should most effective at
the longest azimuthal wavelength.

\begin{figure}
\begin{center}
\includegraphics[width=\columnwidth]{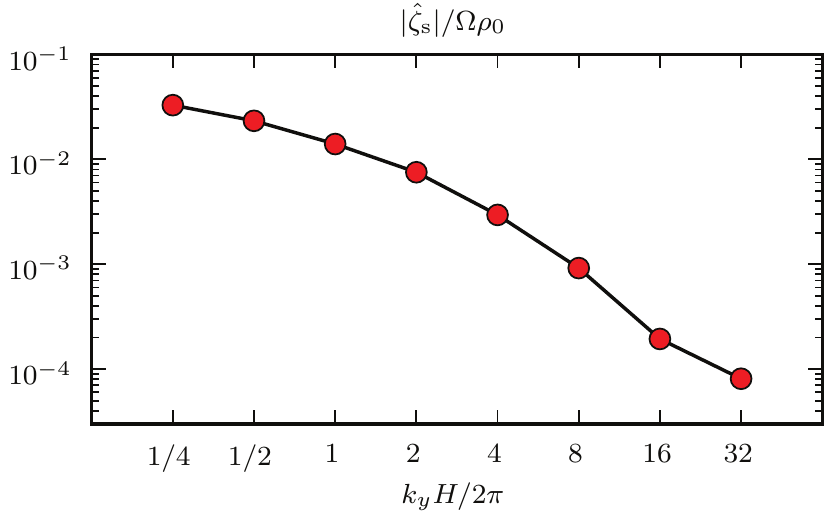}
\end{center}
\caption{The spectrum of swing pseudo potential vorticity averaged over all
swings as a function of azimuthal wave number $k_y$. The spectrum peaks at the
smallest $k_y$, remains rather shallow at moderately small $k_y$, but falls
off for high $k_y$.}
\label{fig:zeta-spectrum}
\end{figure}

The other important dependence is that of the PPV
spectrum on $k_y$ at the time of swing.
This spectrum is difficult to predict theoretically,
but easily determined from the simulation data. We do so by taking
for each $k_y$ an average over the ensemble of shearing waves that we find to
swing from leading to trailing during the course of the simulation. The result
is shown in Fig.~\ref{fig:zeta-spectrum}.
We see that the spectrum is rather shallow on large scales and falls off
at small scales. A nearly flat spectrum at small $k_y$ means that the
dependence of the wave amplitude on $k_y,$ at the most effective values, is
determined by the WKBJ exponential factor appearing in
e.g.\ (\ref{eq:limiting-value-xi}), and should be maximal at either
\mbox{$k_y=1/H$}, or the smallest possible $k_y$ in the box should that be
larger. Accordingly we expect the smallest $k_y$ to be dominant for $L_y=4H$
but this dominance should begin to be lost once $L_y$ starts to exceed
$2\pi{}H.$ We checked that when $L_y$ was extended to $8H,$ the contribution
from the smallest $k_y$ no longer exceeded that from the next smallest
confirming the discussion given above.

Differentially rotating discs undergoing MRI driven turbulence are ubiquitous
in astrophysics and the results presented here and in paper~I indicate that
the production of density waves is generic. The density perturbations
associate with these may play an important role in driving protoplanetary
migration \citep[e.g.][]{2004MNRAS.350..849N,2005A&A...443.1067N} and may play
a role in exciting quasi periodic oscillations in discs around close binary
stars. In the case of protoplanetary discs it has been suggested that only the
high altitude regions may be turbulent while the mid-plane region lies in a
`dead zone' \citep[e.g.][]{1996ApJ...462..725G}. We comment that we found that
the SD waves excited by the turbulence have very weak vertical dependence.
Although we considered an unstratified disc, SD waves with weak vertical
dependence also exist in the isothermal stratified case
\citep[e.g.][]{2007A&A...468....1F}, so that although they may be excited in
the upper layers propagation throughout the vertical structure is to be
expected with the consequence that some small but non zero transport may
result throughout the disc. In this context we note that, consistently with
these ideas, recent simulations by \citet{2007ApJ...670..805O} that
incorporated dead zones were found to have wavelike motions excited in the
mid-plane regions associated with a small residual angular momentum flux that
was uniformly distributed in the dead zone with a typical $\alpha$ value
between $10^{-4}$ and $10^{-5}$. We remark that a simple application of the
WKBJ theory developed in paper~I suggests, since the source of wave
excitation is a vertical average of the pseudo potential vorticity, and the
wave angular momentum flux is the square of this,
that the value of $\alpha$ in the dead zone
should scale as ratio of the square of the surface density in the high
altitude turbulent regions to
the surface density of the whole disc.  However, this may be a lower bound
because the dependence on surface density may actually be weaker on account of
weaker dissipation in the linear regime allowing the build up of a relatively
larger amount of wave angular momentum flux.
This aspect remains to be investigated and
decided by future much higher resolution simulations that we are unable to
perform at the present time.

\section*{Acknowledgements}
T.~H.\ acknowledges support from the Science and Technology Facilities Council
(STFC) as well as from the Isaac Newton Trust. The authors thank Geoffroy
Lesur for fruitful discussions. Computing resources were provided by the
Danish Centre for Scientific Computing (DCSC) and the University of Cambridge
High Performance Computing Service.

\appendix

\section{The Pencil Code}\label{app:pencil-code}
The Pencil Code \citep[see e.g.][]{Brandenburg2002471} is a high-order
finite-difference MHD code that is primarily designed to deal with weakly
compressible turbulent flows. Solenoidality is ensured by solving the
induction equation in terms of the magnetic vector potential. The equations of
motion are evolved using an explicit $3^\mathrm{rd}$ order Runge-Kutta time
stepping scheme. The scheme is stabilized by explicit diffusion coefficients.

For the purposes of the present paper, the shearing box equations that the
code solved numerically consist of the continuity equation
\begin{equation*}
\mathcal{D}\rho + \vec{u}\cdot\nabla\rho = - \rho\nabla\cdot\vec{u}
+ \dot{\rho}_\mathrm{diff},
\end{equation*}
the momentum equation for an isothermal equation of state
\begin{multline*}
\mathcal{D}\vec{u} + \vec{u}\cdot\nabla\vec{u} = -2\vec{\Omega}\times\vec{u}
+ q\Omega u_x\vec{e}_y \\
- \frac{c^2\nabla\rho}{\rho}
+ \frac{\vec{j}\times\vec{B}}{\rho} + \vec{\dot{u}}_\mathrm{diff},
\end{multline*}
and the induction equation
\begin{equation*}
\mathcal{D}\vec{A} = q\Omega A_y\vec{e}_x + \vec{u}\times\vec{B}
+ \vec{\dot{A}}_\mathrm{diff},
\end{equation*}
where
\begin{equation*}
\mathcal{D} = \partial_t - q\Omega x\partial_y,
\end{equation*}
$\rho$ is the mass density, $\vec{u}$ is the fluid velocity, $\vec{A}$ is the
magnetic vector potential, \mbox{$\vec{B}=\nabla\times\vec{A}$} is the
magnetic field, \mbox{$\vec{j}=\nabla\nabla\cdot\vec{A}-\nabla^2\vec{A}$} is
the current density, \mbox{$\vec{\Omega}=\Omega\vec{e}_z$} is the angular
velocity, and $c$ is the speed of sound. For Keplerian rotation $q=3/2$.

The numerical scheme is stabilised by imposing explicit diffusion
terms. For the momentum equation and the induction these take the same form as
molecular diffusion, i.e.
\begin{equation*}
\vec{\dot{u}}_\mathrm{diff} =
\nu\left(\nabla^2\vec{u} + \frac{1}{3}\nabla\nabla\cdot\vec{u}
+ \frac{2\vec{S}\cdot\nabla\rho}{\rho}\right)
\end{equation*}
and
\begin{equation*}
\vec{\dot{A}}_\mathrm{diff} =
\eta\left(\nabla^2\vec{A} - \nabla\nabla\cdot\vec{A}\right),
\end{equation*}
where the traceless rate-of-strain tensor
\begin{equation*}
S_{ij} = \frac{1}{2}\left(\pder{u_i}{x_j} + \pder{u_j}{x_i}
- \frac{2}{3}\delta_{ij}\nabla\cdot\vec{u}\right)
- \frac{q\Omega}{2}(\delta_{ix}\delta_{jy} + \delta_{iy}\delta_{jx}).
\end{equation*}
the values of kinematic viscosity $\nu$ and magnetic resistivity are given in
the text.

In the case of the continuity equation, where there is no explicit physical
diffusion coefficient, we maintain numerical stability by employing
$6^\mathrm{th}$ order hyper diffusion, i.e.
\begin{equation*}
\label{eq:mass-diff}
\dot{\rho}_\mathrm{diff} =
\delta\left(\frac{\partial^6}{\partial x^6}
           +\frac{\partial^6}{\partial y^6}
           +\frac{\partial^6}{\partial z^6}\right)\rho,
\end{equation*}
where the mass diffusion coefficient $\delta$ is chosen to scale with the grid
spacing $\delta x$,
\begin{equation*}
\delta \lesssim \frac{(\delta x)^5 c}{60},
\end{equation*}
so that significant mass diffusion only occurs near the grid scale. We note
that the mass diffusion term (\ref{eq:mass-diff}) formally conserves mass and
that we indeed find very little variation in the total mass during the course
of the simulation.

All derivative operators are given by $6^\mathrm{th}$ order accurate
central finite-difference formulae,
\begin{gather*}
f'_i = \frac{45(f_{i+1}-f_{i-1}) - 9(f_{i+2}-f_{i-2}) + (f_{i+3}-f_{i-3})}
{60\,\delta x}\\
f''_i = \frac{-490 f_i + 270(f_{i+1}+f_{i-1}) - 27(f_{i+2}+f_{i-2})
+ 2(f_{i+3}+f_{i-3})}{180(\delta x)^2}\\
f^{(6)}_i = \frac{20 f_i - 15(f_{i+1}+f_{i-1})
+ 6(f_{i+2}+f_{i-2}) - (f_{i+3}-f_{i-3})}{(\delta x)^6}
\end{gather*}

The shearing periodic boundary condition (\ref{eq:shearing-periodic-bcx}) is
implemented using $6^\mathrm{th}$ order polynomial interpolation.

\section{Discrete Fourier transforms in the shearing sheet}\label{app:fourier}
Let us consider a prototypical evolution equation for a quantity $f(x,y,t)$ in
the shearing sheet of the form
\begin{equation}
\label{eq:sheet-proto}
\left(\partial_t - q\Omega x\partial_y\right)f = c\,\partial_x f
\end{equation}
We may Fourier-decompose this equation in $y$, but because of the explicit
$x$-dependence we are not allowed to do so in $x$. We can however transform to
shearing coordinates,
\begin{align*}
\tilde{x} &= x \\
\tilde{y} &= y + q\Omega t x \\
\tilde{t} &= t
\end{align*}
so that
\begin{align*}
\partial_x &= \partial_{\tilde{x}} + q\Omega t\partial_{\tilde{y}} \\
\partial_y &= \partial_{\tilde{y}} \\
\partial_t &= \partial_{\tilde{t}} + q\Omega x\partial_{\tilde{y}}.
\end{align*}
In shearing coordinates (\ref{eq:sheet-proto}) reads
\begin{equation}
\label{eq:sheet-proto-sheared}
\partial_{\tilde{t}}\tilde{f} =
c(\partial_{\tilde{x}} + q\Omega t\partial_{\tilde{y}})\tilde{f}
\end{equation}
where the transformed field $\tilde{f}$ is defined through
\begin{equation}
\label{eq:transformed-field}
\tilde{f}(\tilde{x},\tilde{y},\tilde{t}) =
\tilde{f}(x,y+ q\Omega t x ,t) = f(x,y,t)
\end{equation}
and so is obtained by shifting $y$ by $q\Omega t x$.
By transforming to shearing coordinates we have eliminated the explicit
$x$-dependence at the expense of an explicit $t$-dependence on the right hand
side of (\ref{eq:sheet-proto-sheared}). We now impose fully periodic boundary
conditions in shearing coordinates, i.e.
\begin{align*}
\tilde{f}(\tilde{x} + L_x,\tilde{y},\tilde{t}) &=
\tilde{f}(\tilde{x},\tilde{y},\tilde{t}) \\
\tilde{f}(\tilde{x},\tilde{y} + L_y,\tilde{t}) &=
\tilde{f}(\tilde{x},\tilde{y},\tilde{t}).
\end{align*}
From (\ref{eq:transformed-field}) it then follows that
\begin{equation*}
f(x + L_x, y - q\Omega t L_x, t) = f(x,y,t)
\end{equation*}
and
\begin{equation*}
f(x, y + L_y, t) = f(x,y,t)
\end{equation*}
which are just the shearing sheet boundary conditions that require periodicity
in shearing coordinates

We are now in a position to Fourier decompose any field $f(x,y,t)$ in shearing
coordinates,
\begin{equation*}
\tilde{f}(\tilde{x},\tilde{y},\tilde{t}) =
\sum_{n_x,n_y}\hat{f}_{n_{x},n_{y}}(\tilde{t})
\exp\left[\rmi\bigg(\!\frac{2\pi n_x}{L_x}\!\bigg)\tilde{x}
         +\rmi\bigg(\!\frac{2\pi n_y}{L_y}\!\bigg)\tilde{y}\right]
\end{equation*}
where the Fourier coefficients can be computed from the inverse transform
\begin{multline*}
\hat{f}_{n_{x},n_{y}}(\tilde{t}) =
\frac{1}{L_x L_y}\iint\!\tilde{f}(\tilde{x},\tilde{y},\tilde{t}) \\
\exp\left[-\rmi\bigg(\!\frac{2\pi n_x}{L_x}\!\bigg)\tilde{x}
          -\rmi\bigg(\!\frac{2\pi n_y}{L_y}\!\bigg)\tilde{y}\right]
\rmd\tilde{x}\,\rmd\tilde{y}.
\end{multline*}
Because $\rmd\tilde{x}\,\rmd\tilde{y}=\rmd{}x\,\rmd{}y$ it follows from
\begin{equation*}
\hat{f}_{n_x,n_y}(t) = \frac{1}{L_x L_y}
\iint\!f(x,y,t)\exp\Big[-\rmi k_x(t) x - \rmi k_y y\Big]\rmd x\,\rmd y
\end{equation*}
where the wave numbers in unsheared coordinates are
\begin{equation*}
k_x(t) = \bigg(\frac{2\pi n_x}{L_x}\bigg)
+ q\Omega t\bigg(\frac{2\pi n_y}{L_y}\bigg)
\quad\mathrm{and}\quad
k_y = \bigg(\frac{2\pi n_y}{L_y}\bigg).
\end{equation*}
More explicitly
\begin{multline*}
\hat{f}_{n_x,n_y} = \frac{1}{L_x}
\int\!\exp\left[-\rmi q\Omega t\bigg(\frac{2\pi n_y}{L_y}\bigg)x\right] \\
\left\{\frac{1}{L_y}\int\!f(x,y,t)\,
\exp\left[\rmi\bigg(\frac{2\pi n_y}{L_y}\bigg)\right]\rmd y\right\}
\exp\left[\rmi\bigg(\frac{2\pi n_x}{L_x}\bigg)\right]\rmd x
\end{multline*}

In order to compute the Fourier transform of a real quantity from the
simulation we therefore do an FFT in $y$ first, multiply the result with
\begin{equation}
\label{eq:phase-shift}
\exp\left[-\rmi q\Omega t\bigg(\frac{2\pi n_y}{L_y}\bigg)x\right],
\end{equation}
and finally do an FFT in $x$. We then determine the time dependent radial wave
number that each Fourier amplitude corresponds to from
\begin{equation*}
k_x(t) = \bigg(\frac{2\pi n_x}{L_x}\bigg)
+ q\Omega t\bigg(\frac{2\pi n_y}{L_y}\bigg)
+ m\bigg(\frac{\pi N_x}{L_x}\bigg)
\end{equation*}
where $N_x$ is the total number of grid points in the $x$-direction and the
integer $m$ is chosen such that we always have
\begin{equation*}
-\bigg(\frac{\pi N_x}{L_x}\bigg) \le k_x(t)
\le \bigg(\frac{\pi N_x}{L_x}\bigg).
\end{equation*}

Note that the phase shift (\ref{eq:phase-shift})
corresponds to an $x$-dependent translation along $y$ by an amount
$-q\Omega t x$ which makes the field fully periodic in $x$.

\bibliographystyle{mn2e}
\bibliography{simulations}

\label{lastpage}

\end{document}